# Construction of the Digital Hadron Calorimeter

Kurt Francis on behalf of the CALICE Collaboration

Argonne National Laboratory, 9700 S. Cass Avenue, Argonne, IL 60439, U.S.A.

Particle Flow Algorithms (PFAs) have been proposed as a method of improving the jet energy resolution of future colliding beam detectors. PFAs require calorimeters with high granularity to enable three-dimensional imaging of events. The Calorimeter for the Linear Collider Collaboration (CALICE) is developing and testing prototypes of such highly segmented calorimeters. In this context, a large prototype of a Digital Hadron Calorimeter (DHCAL) was developed and constructed by a group led by Argonne National Laboratory. The DHCAL consists of 52 layers, instrumented with Resistive Plate Chambers (RPCs) and interleaved with steel absorber plates. The RPCs are read out by 1 x 1 cm$^2$ pads with a 1-bit resolution (digital readout). The DHCAL prototype has approximately 480,000 readout channels. This talk reports on the design, construction and commissioning of the DHCAL. The DHCAL was installed at the Fermilab Test Beam Facility in fall 2010 and data was collected through the summer 2011.

## 1 Introduction

Future particle colliders require a detector with superior jet energy resolution to distinguish W and Z boson hadronic decays on an event-by-event basis. A favored method to improve the jet energy resolution is to apply Particle Flow Algorithms (PFAs) that measure the energy of the various components of a jet with the part of the detector capable of the highest resolution for that component. For example, charged particles are measured with the tracker which has very high momentum resolution, gammas are measured by the electromagnetic calorimeter and neutral hadrons by the combined electromagnetic and hadronic calorimeter. For this technique to work effectively, an excellent tracker in a high magnetic field and a dense calorimeter with high granularity and placed preferably inside the coil are required. The high granularity is necessary to identify the charged and uncharged energy deposits so as to avoid double counting, for instance, of the charged components which have already been measured within the tracker.

In this context a so-called Digital Hadron Calorimeter (DHCAL) using Resistive Plate Chambers (RPCs) as active elements and 1 cm x 1 cm readout pads with a single bit readout is being developed. This program is an integral part of the program of the CALICE collaboration [1], which develops imaging calorimeters for use in future lepton colliders.

This talk reports on the design, construction, quality control, and installation of the DHCAL in the Fermilab test beam [2].

## 2 Brief description of the DHCAL

The DHCAL uses an absorber structures originally developed for the CALICE Analog HCAL and Tail Catcher/Muon Tracker (TCMT) see Fig. 1. The HCAL section consists of 38, 1.74 cm thick steel plates. The TMCT has two sections, a thin section with eight 1.9 cm steel plates and a thick section with 10.2 cm plates.

The active layers to be inserted between the absorber plates, are built out of three 32 cm x 96 cm Resistive Plate Chambers (RPCs), stacked on top of each other and thus creating a 96 x 96 cm$^2$ active surface area. The readout of the chambers is segmented into 1 x 1 cm$^2$ pads each with a single threshold or 1-bit resolution, hence, the denomination of a digital calorimeter. In total 52 layers were instrumented giving close to 480,000 readout channels.

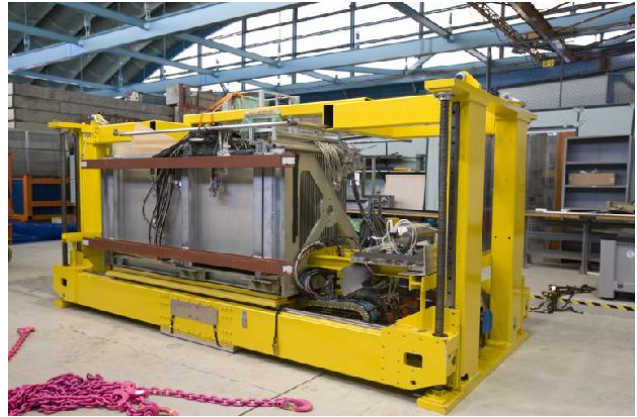

Figure 1: Photograph of the absorber structure and movable stage.

## 3  RPC Design

In a generic RPC, a high voltage (on the order of 6000 to 7000 volts) is applied across a gas gap confined by two resistive plates. RPCs typically use glass or Bakelite as resistive plates. Ionization caused by a charged particle passing through the gap is amplified by the avalanche effect. This signal travels across the resistive plate to the anode where it induces a charge on the readout board. RPCs can operate in either streamer mode or avalanche mode. Avalanche mode is more appropriate for the one bit readout design of the DHCAL but requires greater control of the plate resistivity.

The DHCAL RPC design consists of two glass plates separated by 1.15 mm, see Fig. 2. The thickness of the anode (cathode) glass plates are 0.85 (1.15) mm. The thinner glass on the anode side is to keep the average number of pads with signals above threshold as close as possible to unity for minimum ionizing particles crossing the gas gap. They are glued to a plastic frame and are held apart by fishing lines placed 5 cm apart. The latter also provide a path for the gas to flow uniformly through all parts of the chamber. A unique, 1 – glass RPC has also been developed at Argonne, but was not used for the assembly of the DHCAL. The total RPC thickness is approximately 3.4 mm and has a dead area of 5% including the frame and fishing lines. The following steps are required to produce a RPC: spraying of glass plates with resistive paint, cutting of frame pieces, gluing frame, gluing glass plates onto frame, and mounting of the HV connection. In the following we briefly describe these steps.

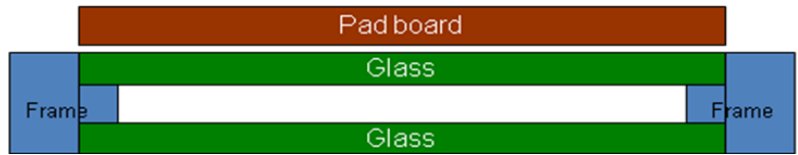

Figure 2: Schematic of the DHCAL RPC design.

## 4 Spraying Glass

To apply a high voltage on the outside of the chamber a paint with a uniform surface resistivity of 1 to 5 M$\Omega$/□ needs to be applied. The lower bounds on the value is defined by the requirement to keep the average pad multiplicity for minimum ionizing particles as close to unity as possible. The upper bound is to avoid efficiency losses at higher particle fluxes. Values on the thicker plate are less stringent.

The required surface resistivity was obtained with a two component 'artist' paint. This non-toxic paint was applied with a spray gun mounted on a computer controlled arm, see Fig. 3. The application took approximately 2 minutes/plane, apart from a significant set-up and cleaning time. After the paint had dried the surface resistivity was measured in a grid over the entire surface of the glass. Figure 4 shows an example of the obtained uniformity. While the uniformity over the surface was in general adequate, large variations of the average value from plate to plate were observed. As a result only 60% of the sprayed plates could be used for constructing RPCs, the others having to be cleaned and re-sprayed.

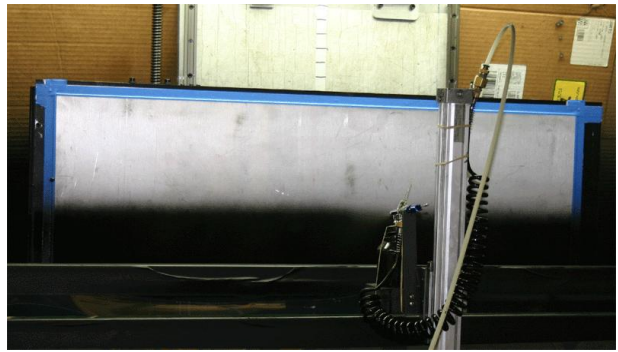

Figure 3: Photograph of the spraying of a sheet of glass with resistive paint.

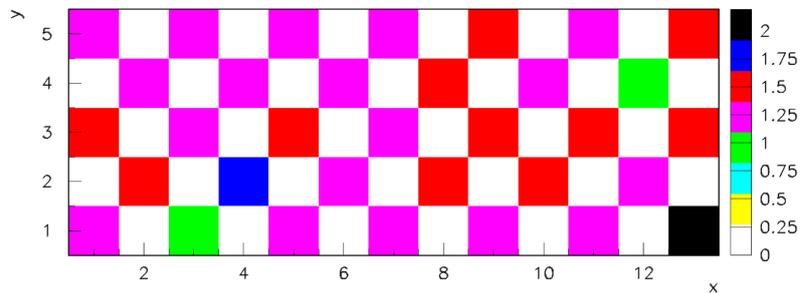

Figure 4: Measured resistivity of a painted glass sample. Values are in MΩ/□. The areas in white were not measured.

## 5 RPC Assembly

Chamber assembly was accomplished in the following sequence. The plastic frame was cut to size and glued together on a custom gluing fixture to provide the correct shape and outer structure. Next a sheet of thick painted glass was placed on the frame and the gaps glued to achieve rigidity and create a seal to contain the gas. After allowing the glue to dry overnight, the frame and glass were flipped over. Four nylon fishing lines, covered to approximately 90% of their length with plastic sleeves (1.15 mm diameter) were inserted with tension across the length of the glass. The sleeves were positioned so as to form a path to direct the gas flow from the input to the output. Next a sheet of thin painted glass was placed over the frame and glued in place. Once the glue was dry, the gas inlet tubes, high voltage mounting fixtures and the fishing lines were glued in place. Finally, the RPCs were covered in insulating mylar and the high voltage cables were attached. A total of 114 (42) chambers were needed for the DHCAL (TCMT) plus spares. All together, 205 RPC's were produced. Figure 5 shows a photograph of one of the three assembly stations including a partially completed chamber with fishing lines/gas spaces installed.

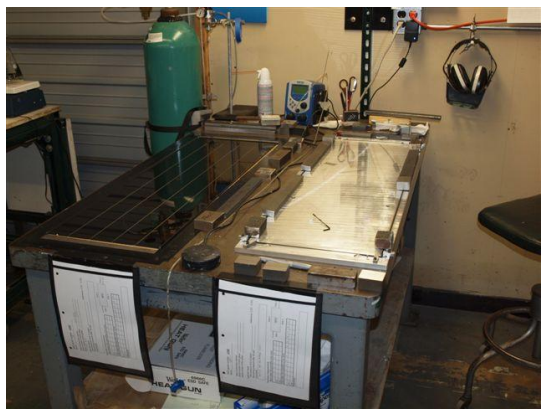

Figure 5: Assembly station for the construction of RPC chambers.

## 6  Quality Assurance

Completed chambers were first given a pressure test with the equivalent of 0.3 inch of water pressure. The chamber passed the test if the pressure drop was 0.02 inch or less in 30 seconds. Chambers not passing the first test were repaired and subsequently retested.

To assure uniformity in the gap thickness, the thickness of a sample of chambers was measured along the edges, see Fig 6. Since the glass is very uniform this measurement provided a measure of the gap size. In general gap sizes at the edges were found to be within 0.1 mm. The central region is assumed to be uniform due to the fishing lines/gas spacers. Corners were typically thicker (up to 0.3 – 0.4 mm) but affecting only a very small region. Five RPC's were found to have low efficiency regions at the corners or along the sides, due to larger gaps at those places. These chambers were eventually removed from the stack and replaced with chambers with more uniform efficiency.

Every chamber was tested at 7.0 kV (the default operating voltage was 6.3 kV) for approximately 24 hours. To be accepted for installation the chamber needed to maintain a current of less than 0.3 uA.

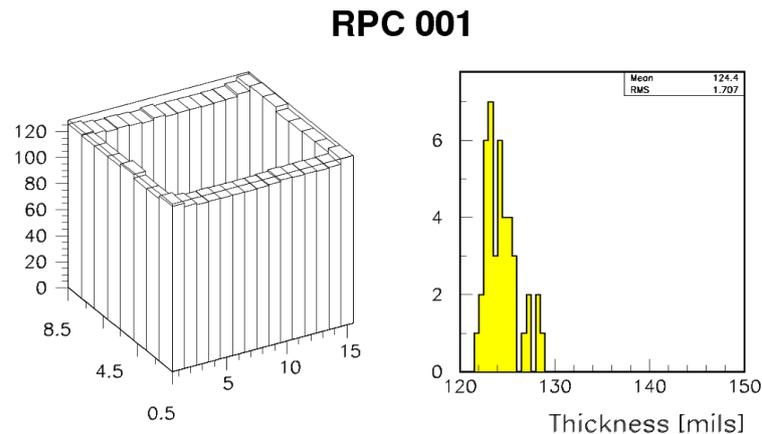

Figure 6: Measurement of edge thickness in mils (equivalent to 25 um).

## 7  Cosmic Ray Test Stand

For testing of early FE board versions and for the first batch of chambers, a cosmic ray test stand was used, see Fig. 7. Up to 9 RPCs could be tested at once. Response to cosmic rays and noise measurements were performed. The chambers were characterized using cosmic rays and noise. High voltage scans were also performed from 5.8 to 6.8 kV to find the optimal HV setting. (6.3kV). Fig. 8 shows the response of an RPC to cosmic rays. The top graph shows the frequency of cosmic ray hits, the middle graph is the average pad multiplicity and the bottom the average efficiency all as a function of angle of incidence. The pad multiplicity and efficiency are both higher at higher angles because the incoming particles traverse more of the gas volume than they do at lower angles.

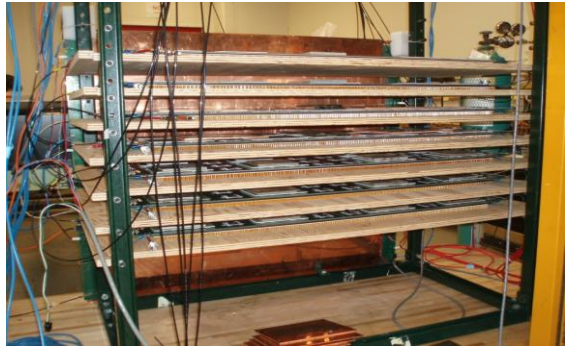

Figure 7: Cosmic ray test stand.

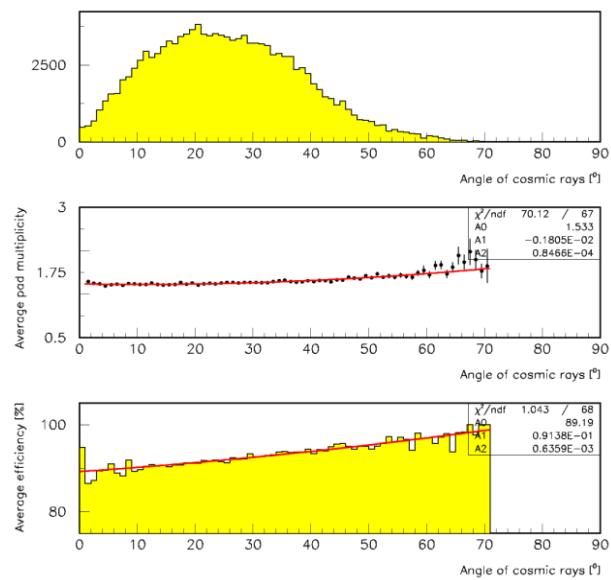

Figure 8: Dependence of RPC response as a function of the angle of incidence of cosmic rays.

# 8  DCAL Chip

The readout of the chambers was performed with a 64-channel ASIC, the DCAL chip developed jointly by FNAL and Argonne, see Fig. 9. It operates in two gain modes: a high gain appropriate for use with GEMs and micromegas with a minimum threshold of approximately 5 fC and a low gain appropriate for use with RPCs with minimum threshold of approximately 30 fC. Each DCAL chip has a threshold that is set by an 8 – bit DAC (up to ~600 fC) that is common to all 64 channels. In addition to a triggered readout mode for cosmic rays and test beam data collection, a triggerless mode is available for noise measurements.

DCAL development went through three versions: DCAL I: initial round (analog circuitry not optimized), DCAL II: some minor problems (used in tests of a small scale prototype with up to 2,000 readout channels), and DCAL III: no identified problems (final production: used in current test beam).

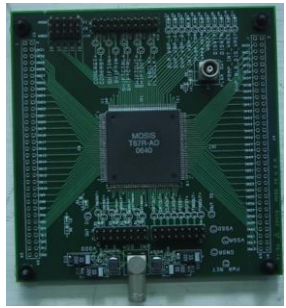

Figure 9: Photograph of the DCAL chip mounted on an evaluation board.

Each RPC is read out by 32 x 96 pads or a total of 3072 pads or channels. Two circuit boards covered the active surface of a chamber, each having 1536 pads and requiring 24 DCAL chips. A data concentrator is implemented into the same board. The front end (FE) board and pad board were manufactured separately to avoid blind and buried vias, a cost and feasibility issue, see Fig. 10.

The pads on the pad board were connected to so-called gluing pads on the other side of the board. These in turn were connected to gluing pads on the Front-end board using conductive epoxy.

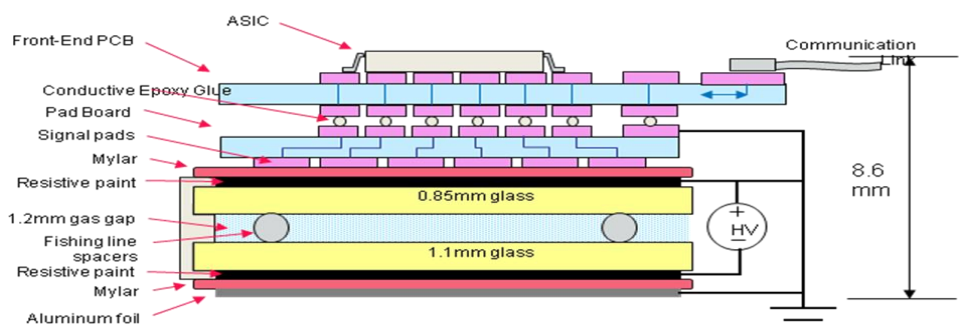

Figure 10: RPC, front end board and pad board configuration. Dimensions not to scale.

The FE board underwent extensive computer-controlled tests including measurements of S-curves and noise rates, etc. These tests required three to six hours per board. Boards with less than 4/1536 dead channels were accepted. Fig. 11 shows a photograph of a completed FE board.

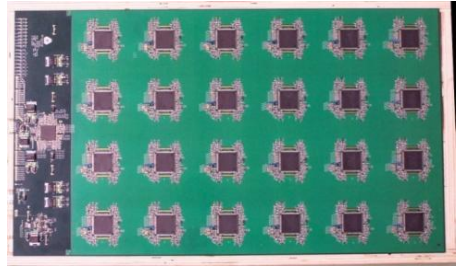

Figure 11: Photograph of front end board. The darker area corresponds to the data concentrator.

## 9 Cassette Assembly

The detector cassette is constructed with a 2 mm steel back, three RPCs, six front end board/ pad board combinations and a 2 mm copper front piece. The cassette is compressed horizontally with a set of 4 (Badminton) strings. The strings are tensioned to ~20 lbs each. This keeps the copper plate in thermal contact with the surface of the DCAL chips for cooling purposes. Each cassette took approximately 45 minutes to assemble. The cassettes were tested with cosmic rays before being shipped to the test beam at Fermilab, Fig. 12.

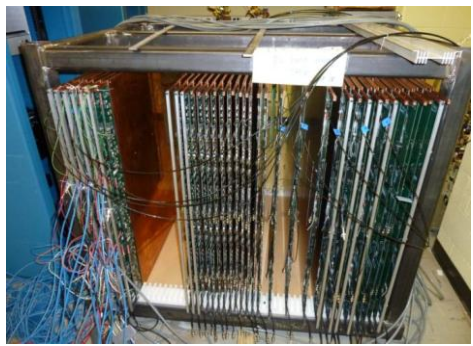

Figure 12: Photograph of completed cassettes undergoing cosmic ray testing.

## 10 Test Beam Installation

The tested cassettes were transported by truck in groups of up to six in a protective shipping fixture to the Fermilab Test Beam Facility where they were installed between the steel absorbers of the HCAL and TCMT stacks. After inserting the cassettes, high voltage, low voltage, and data cables and gas tubes were installed. The first 38 layers took approximately 12 hours to cable up. Fig. 13 shows a photograph of the completed installation at the test beam. The system was first calibrated with muons to characterize the average pad multiplicity and MIP detection efficiency. Pion and positron runs were then performed at energies from 2 GeV to 60 GeV. In addition, data with the 120 GeV primary proton beam were also taken. An example pion event can be seen in Fig. 14.

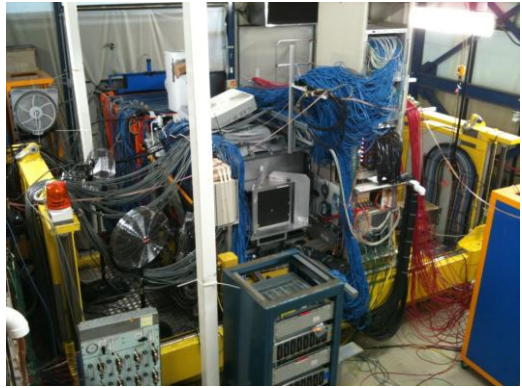

Figure 13: Photograph of DHCAL installation at test beam.

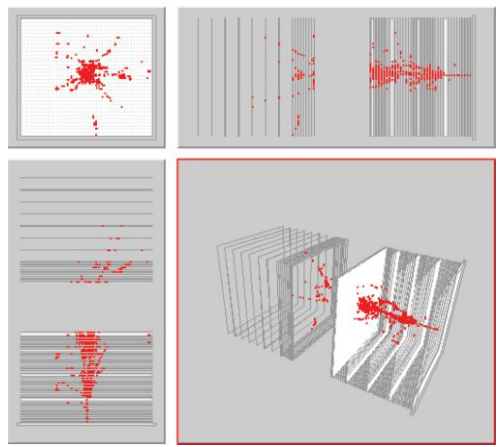

Figure 14: Example of a pion event.

## 11 Summary

The construction of the DHCAL prototype (+TCMT) was completed in February 2011. A series of test beam runs at the Fermilab test beam facility have been very successful and the detector works extremely well. High quality data has been collected and analysis is on-going.

## 12 Acknowledgments

The author would like to thank the organizers for the opportunity to present the DHCAL project.

## 13 References

[1] https://twiki.cern.ch/twiki/bin/view/CALICE/WebHome
[2] http://www-ppd.fnal.gov/FTBF/
[3] J. Repond, Analysis of Muon Events in The DHCAL, these proceedings.
[4] L. Xia, DHCAL Response to Positrons and Pions, these proceedings.